\documentstyle[12pt,fancybox,pst-node]{article}
\title{On the new approach to variable separation
in the time-dependent Schr\"odinger equation with two space
dimensions}
\author{R.Z. ~Zhdanov \\ \small Arnold-Sommerfeld
Institute for Mathematical Physics,\\ \small Leibnitzstra\ss e 10,
38678 Clausthal-Zellerfeld, Germany\thanks{On leave from the Institute
of Mathematics of the Academy of Sciences of Ukraine,
Tereshchenkivska Str.3, 252004 Kiev, Ukraine\newline\indent
e-mail: asrz@pta3.pt.tu-clausthal.de}
\and I.V. ~Revenko and W.I. ~Fushchych \\ \small Institute
of Mathematics of the Academy of Sciences of Ukraine,\\
\small Tereshchenkivska Str.3, 252004 Kiev, Ukraine}
\def\mybox#1{\shadowbox{\begin{tabular}[t]{c} #1 \end{tabular}}}
\let\ds\displaystyle
\begin{document}
\maketitle
\begin{abstract}We suggest an effective approach to separation of
  variables in the Schr\"odinger equation with two space variables.
  Using it we classify inequivalent potentials $V(x_1,x_2)$ such that
  the corresponding Schr\" odinger equations admit separation of
  variables. Besides that, we carry out separation of variables in the
  Schr\" odinger equation with the anisotropic harmonic oscillator
  potential $V=k_1x_1^2+k_2x_2^2$ and obtain a complete list of
  coordinate systems providing its separability. Most of these
  coordinate systems depend essentially on the form of the potential
  and do not provide separation of variables in the free Schr\"
  odinger equation ($V=0$).
\end{abstract}
\section* {I. Introduction}
The problem of separation of variables ({\it SV}) in the two-dimensional
Schr\" odinger equation

\begin{equation}
\imath u_t+u_{x_1x_1}+u_{x_2x_2}=V(x_1,x_2)u
\label{1}
\end{equation}
as well as the most of classical problems of mathematical physics
can be formulated in a very simple way (but this simplicity does not,
of course, imply an existence of easy way to its solution). To
separate variables in Eq.(\ref{1}) one has to construct such functions
$R(t,\vec x), \ \omega _1(t,\vec x), \ \omega _2(t,\vec x)$ that the
Schr\" odinger equation (\ref{1}) after being rewritten in the new
variables

\begin{equation}
\begin{array}{l}
z_0=t, \ z_1=\omega _1(t,\vec x), \ z_2=\omega _2(t,\vec x),\\[1mm]
v(z_0,\vec z)=R(t,\vec x)u(t,\vec x)\\
\end{array}
\label{2}
\end{equation}
separates into three ordinary differential equations ({\it
  ODEs}). From this point of view the problem of {\it SV} in
Eq.(\ref{1}) is studied in \cite{bo1}--\cite{sha}.

But no less important problem is the one of description of potentials
$V(x_1,x_2)$ such that the Schr\" odinger equation admits variable
separation. That is why saying about {\it SV} in Eq.(\ref{1}) we imply
two mutually connected problems. The first one is to describe all such
functions $V(x_1,x_2)$ that the corresponding Schr\" odinger equation
(\ref{1}) can be separated into three {\it ODEs} in some coordinate
system of the form (\ref{2}) (classification problem). The second
problem is to construct for each function $V(x_1,x_2)$ obtained in
this way all coordinate systems (\ref{2}) enabling us to carry out
{\it SV} in Eq.(\ref{1}).

Up to our knowledge, the second problem has been solved provided $V=0$
\cite{bokm,mi} and $V=\alpha x_1^{-2}+\beta x_2^{-2}$\cite{bo1}. The
first one was considered in a restricted sense in \cite{sha}. Authors
using symmetry approach to classification problem obtained some
potentials providing separability of Eq.(\ref{1}) and carried out {\it
  SV} in the corresponding Schr\" odinger equation. But their results
are far from being complete and systematic. The necessary and
sufficient conditions imposed on the potential $V(x_1,x_2)$ by the
requirement that the Schr\" odinger equation admits symmetry operators
of an arbitrary order are obtained in \cite{niof}. But so far there is
no systematic and exhaustive description of potentials $V(x_1,x_2)$
providing {\it SV} in Eq.(\ref{1}).

To have a right to say about description of {\it all} potentials and
{\it all} coordinate systems making it possible to separate the Schr\"
odinger equation one has to give a definition of {\it SV}. One of the
possible definitions of {\it SV} in partial differential equations
({\it PDEs}) is proposed in our paper \cite{fuzr}. It is based on the
concept of Ansatz suggested by Fushchych \cite{fu} and on ideas
contained in the paper by Koornwinder \cite{ko}. The said definition
is quite algorithmic in the sense that it contains a regular algorithm
of variable separation in partial differential equations which can be
easily adapted to handle both linear \cite{fuzr,zrfu} and nonlinear
\cite{zh} {\it PDEs}. In the present paper we apply the said algorithm
to solve the problem of {\it SV} in Eq.(\ref{1}).

Consider the following system of {\it ODEs}:

\begin{eqnarray}
\imath \ds {d\varphi _0\over dt}&=&U_0(t,\varphi _0;\lambda _1,
\lambda _2),\nonumber\\
\ds {d^2\varphi _1\over d\omega _1^2}&=&U_1(\omega _1,
\varphi _1, \ds {d\varphi _1\over d\omega _1};\lambda _1,\lambda
_2),\label{3a}\\
\ds {d^2\varphi _2\over d\omega _2^2}&=&U_2(\omega _2,
\varphi _2, \ds {d\varphi _2\over d\omega _2};
\lambda _1,\lambda _2),\nonumber
\end{eqnarray}
where $U_0, \ U_1, \ U_2$ are some smooth functions of the corresponding
arguments, $\lambda _1, \lambda _2\subset R^1$ are arbitrary
parameters (separation constants) and what is more

\begin{equation}
{\rm rank}\, \left \|\matrix {
\ds {\partial U_{\mu }\over \partial \lambda _a}\\
 }\right \|_{\mu =0\; a=1}^{2\quad\; 2}=2
\label{3b}
\end{equation}
(the last condition ensures essential dependence of the corresponding
solution with separated variables on $\lambda _1, \ \lambda _2$, see
\cite{ko}).

{\bf Definition 1.} We say that Eq.(\ref{1}) admits {\it SV} in the
system of coordinates $t, \ \omega _1(t,\vec x),$ $ \ \omega _2(t,\vec
x)$ if substitution of the Ansatz

\begin{equation}
u=Q(t,\vec x)\varphi _0(t)\varphi _1\Bigl (\omega _1(t,\vec x)
\Bigr)\varphi_2\Bigl (\omega _2(t,\vec x)\Bigr )
\label{4}
\end{equation}
into (\ref{1}) with subsequent exclusion of the derivatives ${d\varphi
_0\over dt}, \ {d^2\varphi _1\over d\omega _1^2}, \ {d^2\varphi _2\over
d\omega _2^2}$ according to equations (\ref{3a}) yields an identity with
respect to $\varphi _0, \ \varphi _1, \ \varphi _2, \ {d\varphi
  _1\over d\omega _1}, {d\varphi _2\over d\omega _2}, \ \lambda _1, \
\lambda _2$.

Thus, according to the above definition to separate variables in
Eq.(\ref{1}) one has

\begin{itemize}
\item{to substitute the expression (\ref{4}) into (\ref{1}),}

\item{to exclude derivatives ${d\varphi_0\over dt}, \ {d^2\varphi
      _1\over d\omega _1^2}, \ {d^2\varphi _2\over d\omega _2^2}$ with
    the help of equations (\ref{3a}),}

\item{to split the obtained equality with respect to the variables
    $\varphi _0, \ \varphi _1, \ \varphi _2, \ {d\varphi _1\over
      d\omega _1}, \ {d\varphi _2\over d\omega _2}, \ \lambda _1,$ $ \
    \lambda _2$ considered as independent.}
\end{itemize}

As a result one gets some over-determined system of {\it PDEs} for the
functions $Q(t,\vec x),$ \ $\omega _1(t,\vec x),$ $\ \omega _2(t,\vec
x)$.  On solving it one obtains a complete description of all
coordinate systems and potentials providing {\it SV} in the Schr\"
odinger equation.  Naturally, an expression {\it complete description}
makes sense only within the framework of our definition. So if one
uses a more general definition it may be possible to construct new
coordinate systems and potentials providing separability of
Eq.(\ref{1}). But all solutions of the Schr\" odinger equation with
separated variables known to us fit into the scheme suggested by us
and can be obtained in the above described way.

\section* {II. Classification of potentials $V(x _1,x _2)$.}

We do not adduce in full detail computations needed because
they are very cumbersome. We shall restrict ourselves to
pointing out main steps of the realization of the above
suggested algorithm.

First of all we make a remark, which makes life a little
bit easier. It is readily seen that a substitution of the form

\begin{eqnarray}
Q&\to &Q^\prime =Q\Psi _1(\omega _1)\Psi _2(\omega _2),\nonumber\\
\omega _a& \to &\omega _a^\prime =\Omega _a(\omega _a),\; a=1,2,\label{5}\\
\lambda _a&\to &\lambda _a^\prime =\Lambda _a(\lambda _1,
\lambda _2),\ a=1,2,\nonumber
\end{eqnarray}
does not alter the structure of relations (\ref{3a}), (\ref{3b}), (\ref{4}).
That is why, we can introduce the following equivalence relation:

$$
(\omega _1, \; \omega _2, \; Q) \ \sim \ (\omega ^\prime _1, \;
\omega ^\prime _2, \; Q^\prime )
$$
provided (\ref{5}) holds with some $\Psi _a, \ \Omega _a, \ \Lambda _a$.

Substituting (\ref{4}) into (\ref{1}) and excluding the derivatives
${d\varphi _0\over dt}, \ {d^2\varphi _1\over d\omega _1^2},
\ {d^2\varphi _2\over d\omega _2^2}$ with the use of equations
(\ref{3a}) we get

\begin{eqnarray*}
& &\imath (Q_t\varphi _0\varphi _1\varphi _2
+QU _0\varphi _1\varphi _2
+Q\omega _{1t}\varphi _0\dot \varphi _1\varphi _2
+Q\omega _{2t}\varphi _0\varphi _1\dot \varphi _2)
+(\triangle Q)\varphi _0\varphi _1\varphi _2\\
& &\quad +2Q _{x_a}\omega _{1x_a}\varphi _0\dot \varphi _1\varphi _2
+2Q _{x_a}\omega _{2x_a}\varphi _0\varphi _1\dot \varphi _2
+Q\Bigl ((\triangle \omega _1)\varphi _0\dot \varphi _1\varphi _2
+(\triangle \omega _2)\varphi _0\varphi _1\dot \varphi _2\\
& &\quad +\omega _{1x_a}\omega_ {1x_a}\varphi _0U_1\varphi _2
+\omega _{2x_a}\omega_ {2x_a}\varphi _0\varphi _1U_2
+2\omega _{1x_a}\omega_ {2x_a}\varphi _0\dot \varphi _1\dot
\varphi _2\Bigr )
=VQ\varphi _0\varphi _1\varphi _2,
\end{eqnarray*}
where the summation over the repeated index $a$ from 1 to 2 is
understood. Hereafter an overdot means differentiation with respect
to a corresponding argument and $\triangle =\partial _{x_1}^2+
\partial _{x_2}^2$.

Splitting the equality obtained with respect to independent
variables $\varphi _1, \ \varphi _2, \ {d\varphi _1\over d\omega _1},
\ {d\varphi _2\over d\omega _2},$ $\ \lambda _1, \ \lambda _2$ we
conclude that {\it ODEs} (\ref{3a}) are linear and up to the
equivalence relation (\ref{5}) can be written in the form

\begin{eqnarray*}
\imath \ds {d\varphi _0\over dt}&=&\Bigl (\lambda _1R_ 1(t)+
\lambda _2R_ 2(t) +R_ 0(t)\Bigr )\varphi _0,\\
\ds {d^2\varphi _1\over d\omega _1^2}&=&\Bigl (\lambda _1B_{11}(\omega _1)
+\lambda _2B_{12}(\omega _1)+B_{01}(\omega _1)\Bigr )\varphi _1,\\
\ds {d^2\varphi _2\over d\omega _2^2}&=&\Bigl (\lambda _1B_{21}(\omega _2)
+\lambda _2B_{22}(\omega _2)+B_{02}(\omega _2)\Bigr )\varphi _2
\end{eqnarray*}
and what is more, functions $\omega _1, \ \omega _2, \ Q$ satisfy an
over-determined system of nonlinear {\it PDEs}

\begin{eqnarray}
&1.&\omega _{1x_b}\omega _{2x_b}=0,\nonumber\\[2mm]
&2.&B_{1a}(\omega _1)\omega _{1x_b}\omega _{1x_b}+
B_{2a}(\omega _2)\omega _{2x_b}\omega _{2x_b}+R_a(t)=0,\;
\;  a=1,2, \nonumber\\[2mm]
&3.&2\omega _{ax_b}Q_{x_b}+Q(\imath \omega _{at}+\triangle \omega _a),
\; \; a=1,2,\label{6}\\[2mm]
&4.&\Bigl (B_{01}(\omega _1)\omega _{1x_b}\omega _{1x_b}+
B_{02}(\omega _1)\omega _{2x_b}\omega _{2x_b}\Bigr )Q+
\imath Q_t+\triangle Q+R_0(t)Q\nonumber\\[2mm]
& &-V(x _1,x _2)Q=0.\nonumber
\end{eqnarray}

Thus, to solve the problem of {\it SV} for the linear Schr\" odinger
equation it is necessary to construct general solution of
system of nonlinear {\it PDEs} (\ref{6}). Roughly speaking, to solve
a linear equation one has to solve a system of {\it nonlinear
equations}! This is the reason why so far there is no complete
description of all coordinate systems providing separability
of the four-dimensional wave equation \cite{mi}.

But in the case involved we have succeeded in integrating
system of nonlinear {\it PDEs} (\ref{6}). Our approach to integration
of it is based on the following change of variables (hodograph
transformation)

\begin{eqnarray*}
& &z_0=t, \ z_1=Z_1(t,\omega _1,\omega _2), \
z_2=Z_2(t,\omega _1,\omega _2),\\
& &v_1=x_1, \ v_2=x_2
\end{eqnarray*}
where $z_0, \ z_1, \ z_2$ are new independent and $v_1, \ v_2$ are new
dependent variables correspondingly.

Using the hodograph transformation determined above we have
constructed the general solution of equations 1--3 from (\ref{6}). It is
given up to the equivalence relation (\ref{5}) by one of the following
formulae:
\begin{eqnarray}
&1.&\omega _1=A(t)x_1+W_1(t), \quad \omega _2=B(t)x_2+W_2(t),\nonumber\\
& &Q(t,\vec x)=\exp \Biggl \{-\ds {\imath \over 4}
\Bigl (\ds {\dot A\over A}x_1^2+
\ds {\dot B\over B}x_2^2\Bigr )-{\imath \over 2}\Bigl
(\ds {\dot W_1\over A}x_1+
\ds {\dot W_2\over B}x_2\Bigr )\Biggr \};\nonumber\\
&2.&\omega _1=\ds {1\over 2}\ln (x_1^2+x_2^2)+W(t), \quad
\omega _2=\arctan \ds {x_1\over x_2},\nonumber\\
& &Q(t,\vec x)=\exp \Biggl \{-\ds {\imath \dot W\over 4}
(x_1^2+x_2^2)\Biggr \};\nonumber\\
&3.&x_1=\ds {1\over 2}W(t)(\omega _1^2-\omega _2^2)+W_1(t), \quad
x_2=W(t)\omega _1\omega _2+W_2(t),\label{7}\\
& &Q(t,\vec x)=\exp \Biggl \{\ds {\imath \dot W\over 4W}\Bigl ((x_1-W_1)^2
+(x_2-W_2)^2\Bigr )+{\imath \over 2}(\dot W_1x_1+\dot W_2x_2)\Biggr
\};\nonumber\\
&4.&x_1=W(t)\cosh \omega _1\cos \omega _2+W_1(t), \quad
x_2=W(t)\sinh \omega _1\sin \omega _2+W_2(t),\nonumber\\
& &Q(t,\vec x)=\exp \Biggl \{\ds {\imath \dot W\over 4W}\Bigl ((x_1-W_1)^2
+(x_2-W_2)^2\Bigr )+{\imath \over 2}(\dot W_1x_1+\dot W_2x_2)\Biggr
\};\nonumber
\end{eqnarray}

Here $A, \ B, \ W, \ W_1, \ W_2$ are arbitrary smooth functions on $t$.

Substituting the obtained expressions for the functions $Q, \ \omega
_1, \ \omega _2$ into the last equation from the system (\ref{6}) and
splitting with respect to variables $x_1, \ x_2$ we get explicit forms
of potentials $V(x_1, x_2)$ and systems of nonlinear {\it ODEs} for
unknown functions $A(t), \ B(t), \ W(t), \ W_1(t), \ W_2(t)$. We have
succeeded in integrating these and in constructing all coordinate
systems providing {\it SV} in the initial equation (\ref{1}).

Here we consider in detail integration of the fourth equation of
system (\ref{6}) for the case 2 from (\ref{7}), since computations
needed are not so lengthy as for other cases.

First, we make several important remarks which introduce an
equivalence relation on the set of potentials $V(x_1,x_2)$.

{\it Remark 1.} The Schr\"odinger equation with the potential

\begin{equation}
V(x_1, x_2)=k_1x_1+k_2x_2 + k_3+V_1(k_2x_1-k_1x_2),
\label{8*}
\end{equation}
where $k_1, \ k_2, \ k_3$ are constants, is transformed to
the Schr\"odinger equation with the potential

\begin{equation}
V^\prime(x^{\prime}_1, x^{\prime}_2)=V_1
(k_2x^{\prime}_1-k_1x^{\prime}_2)
\label{9*}
\end{equation}
by the following change of variables:

\begin{equation}
\begin{array}{l}
t^\prime =t, \quad \vec x^\prime =\vec x+t^2\vec k, \\[2mm]
u^\prime =u\exp \Bigl \{\ds{\imath\over 3}(k_1^2+k_2^2)t^3
+\imath t(k_1x_1+k_2x_2)+\imath k_3t\Bigr \}.
\end{array}
\end{equation}

It is readily seen that the class of Ans\"atze is transformed into
itself by the above change of variables. That is why, potentials
(\ref{8*}) and (\ref{9*}) are considered as equivalent.

{\it Remark 2.} The Schr\"odinger equation with the potential

\begin{equation}
V(x_1, x_2)=k(x_1^2+x_2^2) + V_1\biggl
({x_1\over x_2}\biggr )(x_1^2+x_2^2)^{-1}
\label{8}
\end{equation}
with $k=const$ is reduced to the Schr\" odinger equation with the
potential

\begin{equation}
V^\prime (x_1, x_2)=V _1\biggl ({x^\prime _1\over x^\prime _2}\biggr )
(x_1^{\prime 2}+x_1^{\prime 2})^{-1}
\label{9}
\end{equation}
by the change of variables

$$
t^\prime =\alpha (t), \quad \vec x^\prime =\beta (t)\vec x, \quad
u^\prime =u\exp \Bigl \{\imath \gamma (t)(x_1^2+x_2^2)+\delta (t)\Bigr \},
$$
where $\Bigl (\alpha (t), \; \beta (t), \; \gamma (t), \; \delta
(t)\Bigr )$ is an arbitrary solution of the system of {\it ODEs}

\begin{eqnarray*}
& &\dot \gamma -4\gamma ^2=k, \quad \dot \beta -4\gamma \beta=0,
\quad \dot \alpha -\beta ^2=0, \quad \dot \delta +4\gamma =0
\end{eqnarray*}
such that $\beta \ne 0$.

Since the above change of variables does not alter the structure of
the Ansatz (\ref{4}), when classifying potentials $V(x_1, x_2)$
providing separability of the corresponding Schr\"odinger equation we
consider potentials (\ref{8}), (\ref{9}) as equivalent.

{\it Remark 3.} It is well-known (see e.g. \cite{bo2,ni}) that the
general form of the invariance group admitted by Eq.(\ref{1}) is as
follows

\begin{eqnarray*}
& &t^\prime =F(t,\vec \theta ), \quad x_a^\prime =g_a(t, \vec x, \vec
\theta ), \quad a=1,2,\\
& &u^\prime =h(t, \vec x, \vec \theta )u
+U(t,\vec x),
\end{eqnarray*}
where $\vec \theta =(\theta _1, \theta _2, \dots ,\theta _n)$ are
group parameters and $U(t,\vec x$) is an arbitrary solution of
Eq.(\ref{1}).

The above transformations also do not alter the structure of the
Ansatz (\ref{4}). That is why, systems of coordinates $t ^\prime , \
x_1^\prime , \ x_2^\prime $ and $t, \ x_1, \ x_2$ are considered as
equivalent.

Now we turn to the integration of the fourth equation of system
(\ref{6}).  Substituting into it the expressions for the functions
$\omega_1, \ \omega_2, \ Q$ given by formulae 2 from (\ref{7}) we get

\begin{equation}
\begin{array}{rcl}
V(x_1,x_2)&=&\Bigl (B_{01}(\omega_1)+B_{02}(\omega_2)\Bigr )
\exp \{-2(\omega_1-W)\}+\ds{1\over 4}(\ddot W
-\dot W^2) \\[2mm]
& &\times\exp \{2(\omega_1-W)\}+R_0(t)-\imath\dot W.
\label{10*}
\end{array}
\end{equation}

In the above equality $B_{01}, \ B_{02}, \ R_0(t), \ W(t)$ are unknown
functions to be determined from the requirement that the right-hand
side of (\ref{10*}) does not depend on $t$.

Differentiating (\ref{10*}) with respect to $t$ and taking into
account the equalities
$$\omega_{1 t}=\dot W, \quad \omega_{2 t}=0
$$
we have

\begin{equation}
\dot W\exp\{-2(\omega_1-W)\}\dot B_{01}+\dot \alpha(t)
\exp\{2(\omega_1-W)\}+\dot \beta(t)=0,
\label{11*}
\end{equation}
where $\alpha(t)={1\over 4}(\ddot W-\dot W^2), \ \beta(t)=R_0-\imath
\dot W$.

Cases $\dot W=0$ and $\dot W\ne 0$ have to be considered separately.

{\it The case 1}. $\dot W=0$.

In this case $W=C=const, \ R_0=0$. Since coordinate systems $\omega_1, \
\omega_2$ and $\omega_1+C_1, \ \omega_2+C_2$ are equivalent with
arbitrary constants $C_1, \ C_2$, choosing $C_1=-C, \ C_2=0$ we can
put $C=0$. Hence it immediately follows that

$$
V(x_1,x_2)=\biggl [B_{01}\biggl (\ds{1\over 2}
\ln \, (x_1^2+x_2^2)\biggr )+B_{02}\biggl (\arctan \,
\ds{x_1\over x_2}\biggr )\biggr ](x_1^2+x_2^2)^{-1},
$$
where $B_{01}, \ B_{02}$ are arbitrary functions. And what is more,
the Schr\"odinger equation (\ref{1}) with such potential separates
only in one coordinate system

\begin{equation}
\omega_1={1\over 2}\ln \, (x_1^2+x_2^2), \quad \omega_2=\arctan
\,\ds{x_1\over x_2}.
\label{12*}
\end{equation}

{\it The case 2}. $\dot W\ne 0$

Dividing (\ref{10*}) into $\dot W\exp\{-2(\omega_1-W)\}$ and
differentiating the equality obtained with respect to $t$ we get

$$
\exp\{4\omega_1\}{d\over dt}\Bigl (\dot\alpha(\dot
  W)^{-1}\exp\{-4W\}\Bigr )+\exp\{2\omega_1\}{d\over dt}\Bigl
  (\dot\beta (\dot W)^{-1}\exp\{-2W\}\Bigr )=0,
$$
whence

$$
{d\over dt}\Bigl (\dot\alpha (\dot W)^{-1}\exp\{-4W\}\Bigr )=0, \quad
{d\over dt}\Bigl (\dot\beta (\dot W)^{-1}\exp\{-2W\}\Bigr )=0.
$$

Integration of the above {\it ODEs} yields the following result:

$$
\alpha=C_1\exp\{4W\}+C_2, \quad \beta=C_3\exp\{2W\}+C_4,
$$
where $C_\jmath, \ \jmath=\overline{1,4}$ are arbitrary real
constants.

Inserting the result obtained into (\ref{11*}) we get an equation
for $B_{01}$

$$
\dot B_{01}=-4C_1\exp\{4\omega_1\}-2C_3\exp\{2\omega_1\},
$$
which general solution reads

$$
B_{01}=-C_1\exp\{4\omega_1\}-C_3\exp\{2\omega_1\}+C_5.
$$

In the above equality $C_5$ is an arbitrary real constant.

Substituting the expressions for $\alpha, \ \beta, \ B_{01}$ into
Eq.(\ref{10*}) we have the explicit form of the potential $V(x_1,x_2)$

$$
V(x_1,x_2)=\biggl [B_{02}\biggl (\arctan \,\ds{x_1\over x_2}\biggr
)+C_5\biggr ](x_1^2+x_2^2)^{-1} +C_2(x_1^2+x_2^2)+C_4,
$$
where $B_{02}$ is an arbitrary function.

By force of the Remarks 1, 2 we can choose $C_2=C_4=0$. Furthermore,
due to arbitrariness of the function $B_{02}$ we can put $C_5=0$.

Thus, the case $\dot W\ne 0$ leads to the following potential:

\begin{equation}
V(x_1,x_2)=B_{02}\biggl (\arctan \,\ds{x_1\over x_2}\biggr )
(x_1^2+x_2^2)^{-1}.
\label{13*}
\end{equation}

Substitution of the above expression into Eq.(\ref{10*}) yields
second-order nonlinear {\it ODE} for the function $W=W(t)$

\begin{equation}
\ddot W-\dot W^2=4C_1\exp \{4W\},
\label{14*}
\end{equation}
while the function $R_0$ is given by the formula

$$
R_0=\imath \dot W+C_3\exp \{2W\}.
$$

Integration of {\it ODE} (\ref{14*}) is considered in detail in the
Appendix 1. Its general solution has the form

\noindent
under $C_1\ne 0$

$$
W=-\ds{1\over 2}\ln \, \Bigl ((at-b)^2-4C_1\Bigr )
+\ds{1\over 2}\ln \, a,
$$

\noindent
under $C_1=0$

$$
W=a-\ln \, (t+b).
$$

Substituting obtained expressions for $W$ into formulae 2 from
(\ref{7}) and taking into account the Remark 3 we arrive at the
conclusion that the Schr\"odinger equation (\ref{1}) with the
potential (\ref{13*}) admits {\it SV} in two coordinate systems. One
of them is the polar coordinate system (\ref{12*}) and another one is
the following:

\begin{equation}
\label{15*}
\omega_1={1\over 2}\ln \, (x_1^2+x_2^2)-{1\over 2}\ln \, (t^2\pm 1), \quad
\omega_2=\arctan \,\ds{x_1\over x_2}.
\end{equation}

Consequently, the case 2 from (\ref{7}) gives rise to two classes of
the separable Schr\"odinger equations (\ref{1}).

Cases 1, 3, 4 from (\ref{7}) are considered in an analogous way but
computations involved are much more cumbersome. As a result, we obtain
the following list of inequivalent potentials $V(x_1,x_2)$ providing
separability of the Schr\"odinger equation.

\begin{enumerate}
\item{$V(x_1, x_2)=V_1(x_1)+V_2(x_2);$}
\begin{enumerate}
\item{$V(x_1, x_2)=k_1x_1^2+k_2x_1^{-2}+V_2(x_2), \quad k_2\ne
    0;$}
\begin{enumerate}
\item{$V(x_1,x_2)=k_1x_1^2+k_2x_2^2+k_3x_1^{-2}+k_4x_2^{-2}, \quad
    k_3k_4\ne 0,$}
\item[{}]{$k_1^2+k_2^2\ne 0, \ k_1\ne k_2;$}
\item{$V(x_1, x_2)=k_1x_1^2+k_2x_1^{-2}, \quad k_1k_2\ne
    0;$}
\item{$V(x_1, x_2)=k_1x_1^{-2}+k_2x_2^{-2};$}
\end{enumerate}
\end{enumerate}
\begin{enumerate}
\item[{(b)}]{$V(x_1, x_2)=k_1x_1^2+V_2(x_2);$}
\begin{enumerate}
\item{$V(x_1, x_2)=k_1x_1^2+k_2x_2^2+k_3x_2^{-2}, \quad
    k_1k_3\ne 0, \ k_1\ne k_2; $}
\item{$V(x_1, x_2)=k_1x_1^2+k_2x_2^2, \quad k_1k_2\ne 0, \
    k_1\ne k_2;$}
\item{$V(x_1, x_2)=k_1x_1^2+k_2x_2^{-2}, \quad k_1\ne 0;$}
\end{enumerate}
\end{enumerate}
\item{$V(x_1, x_2)=V_1(x_1^2+x_2^2)+V_2\biggl (\ds {x_1\over
      x_2}\biggr ) (x_1^2+x_2^2)^{-1};$}
\begin{enumerate}
\item{$V(x_1, x_2)=V_2\biggl (\ds {x_1\over x_2}\biggr
    )(x_1^2+x_2^2)^{-1};$}
\item{$V(x_1, x_2)=k_1(x_1^2+x_2^2)^{-1/2}, \quad k_1\ne 0;$}
\end{enumerate}
\item{$V(x_1,x_2)=\Bigl (V_1(\omega _1)+V_2(\omega _2)\Bigr )
(\omega _1^2+\omega _2^2)^{-1},$}
\item[{}]{${\rm where} \quad \omega _1^2-\omega _2^2=2x_1, \
\omega _1\omega _2=x_2;$}
\item{$V(x_1, x_2)=\Bigl (V_1(\omega _1)+V_2(\omega _2)\Bigr )
(\sinh ^2\omega _1+\sin ^2\omega _2)^{-1},$}
\item[{}]{${\rm where} \quad \cosh \omega _1\cos \omega _2=x_1, \
\sinh \omega _1\sin \omega _2=x_2;$}
\item{$V(x_1, x_2)=0.$}
\end{enumerate}

In the above formulae $V_1, \ V_2$ are arbitrary smooth functions,
$k_1, \ k_2, \ k_3, \ k_4$ are arbitrary constants.

It should be emphasized that the above potentials are not inequivalent
in a usual sense. These potentials differ from each other by the fact
that the coordinate systems providing separability of the
corresponding Schr\" odinger equations are different. As an
illustration, we give the following figure:

\begin{center}
\psset{unit=3cm}
\begin{pspicture}(0.4,0)(4.2,6)
  \psset{arrows=->, nodesep=3pt}
  \rput(2,5.5){\rnode{A}{\mybox{
    $V(x_1,x_2)$}}}
  \rput(0.5,4.5){\rnode{B1}{\mybox{
    $V^{(1)}$}}}
  \rput(1.5,4.5){\rnode{B2}{\mybox{
    $V^{(2)}$}}}
  \rput(2.5,4.5){\rnode{B3}{\mybox{
    $V^{(3)}$}}}
  \rput(3.5,4.5){\rnode{B4}{\mybox{
    $V^{(4)}$}}}
  \rput(0.6,3.5){\rnode{C1}{\mybox{
    $k_1x_1^2+k_2x_1^{-2}+V_2(x_2)$}}}
  \rput(2.1,3.5){\rnode{C2}{\mybox{
    $k_1x_1^2+V_2(x_2)$}}}
  \rput(3.15,3.5){\rnode{C3}{\mybox{
    $V_2({x_1\over x_2})r^{-2}$}}}
  \rput(4.1,3.5){\rnode{C4}{\mybox{
    $k_1r^{-1}$}}}
  \rput(1.2,2.5){\rnode{D1}{\mybox{
    $k_1x_1^2+k_2x_2^2+k_3x_1^{-2}+k_4x_2^{-2}$}}}
  \rput(3.3,2.5){\rnode{D2}{\mybox{
    $k_1x_1^2+k_2x_2^2+k_3x_2^{-2}$}}}
  \rput(0.4,1.5){\rnode{E1}{\mybox{
    $k_1x_1^2+k_2x_1^{-2}$}}}
  \rput(1.65,1.5){\rnode{E2}{\mybox{
    $k_1x_1^{-2}+k_2x_2^{-2}$}}}
  \rput(2.8,1.5){\rnode{E3}{\mybox{
    $k_1x_1^2+k_2x_2^2$}}}
  \rput(3.95,1.5){\rnode{E4}{\mybox{
    $k_1x_1^2+k_2x^{-2}$}}}
  \rput(2.2,0.5){\rnode{F}{\mybox{$V=0$}}}
  \ncline{A}{B1}
  \ncline{A}{B2}
  \ncline{A}{B3}
  \ncline{A}{B4}
  \ncline{B1}{C1}
  \ncline{B1}{C2}
  \ncline{B2}{C3}
  \ncline{B2}{C4}
  \ncline{B3}{C4}
  \ncline{C1}{D1}
  \ncline{C2}{D2}
  \ncline{D1}{E1}
  \ncline{D1}{E2}
  \ncline{D2}{E3}
  \ncline{D2}{E4}
  \ncline{E1}{F}
  \ncline{E2}{F}
  \ncline{E3}{F}
  \ncline{E4}{F}
\end{pspicture}
\end{center}

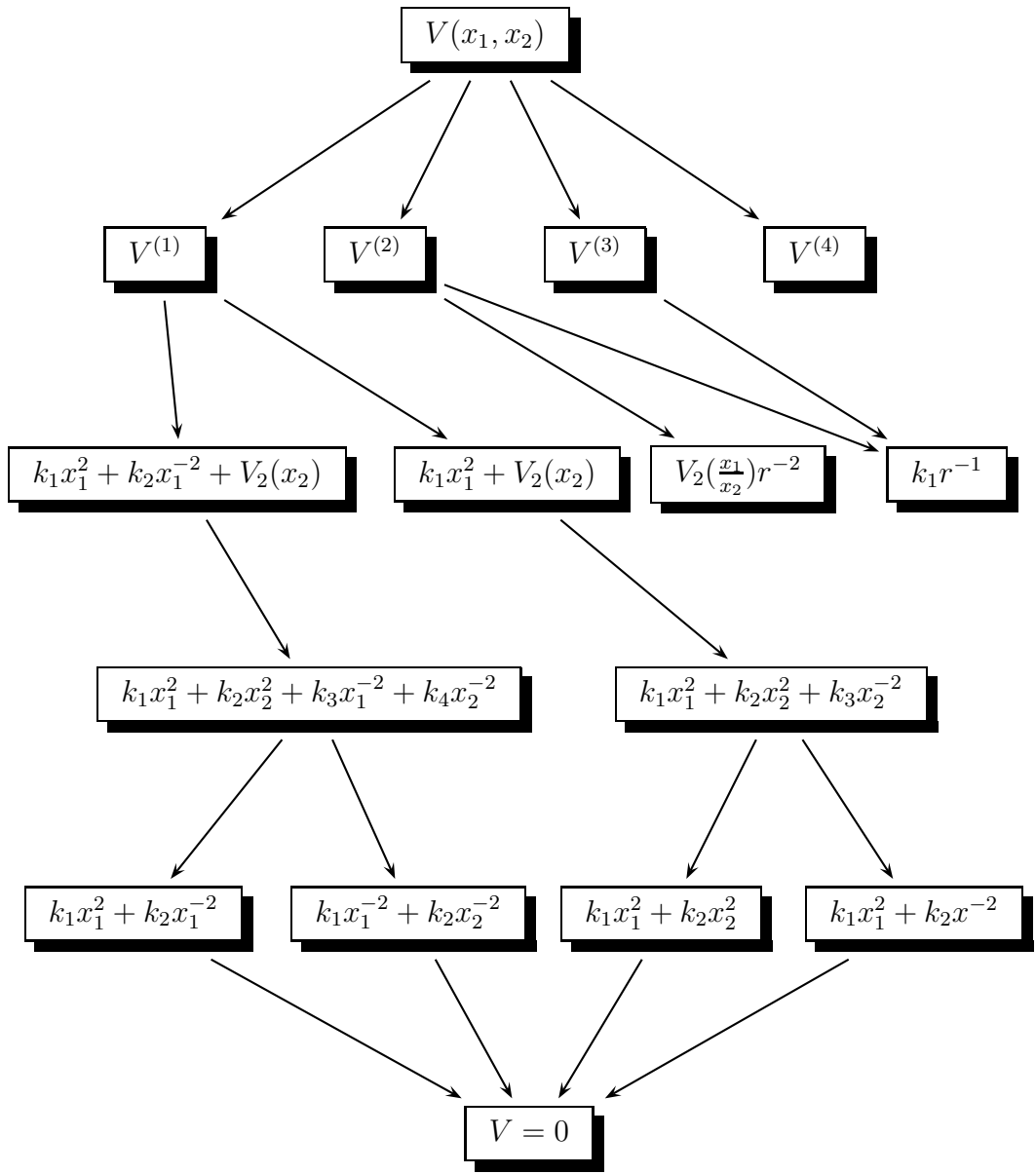
\underline{\bf Fig.1}

\noindent
where $r=(x_1^2+x_2^2)^{1/2}$ and by the symbol $V^{(\jmath)},
\jmath=1,4$ we denote the potential given in the above list under the
number $\jmath$. Downarrows in the Fig.1 indicate specifications of
the potential $Vx_1, x_2)$ providing new possibilities to separate
the corresponding Schr\"odinger equation (\ref{1}).

The Schr\"odinger equation (\ref{1}) with arbitrary function
$V(x_1,x_2)$ (level 1 of the Fig.1) admits no separation of variables.
Next, Eq.(\ref{1}) with the ``root'' potentials $V^{(\jmath)}$ (level
2), $V_1,\ V_2$ being arbitrary smooth functions, separates in the
Cartesian ($\jmath=1$), polar ($\jmath=2$), parabolic ($\jmath=3$) and
elliptic ($\jmath=4$) coordinate systems, correspondingly. Specifying
the functions $V_1,\ V_2$ (i.e. going down to the lower levels) new
possibilities to separate variables in the Schr\"odinger equation
(\ref{1}) arise. For example, Eq.(\ref{1}) with the potential
$V_2({x_1\over x_2})r^{-2}$, which is a particular case of the
potential $V^{(2)}$, separates not only in the polar coordinate system
(\ref{12*}) but also in the coordinate systems (\ref{15*}). The
Schr\"odinger equation with the Coulomb potential $k_1r^{-1}$, which
is a particular case of the potentials $V^{(2)},\ V^{(3)}$, separates
in two coordinate systems (namely, in the polar and parabolic
coordinate systems, see below the Theorem 4). An another
characteristic example is a transition from the potential $V^{(1)}$ to
the potential $k_1x_1^2+V_2(x_2)$. The Schr\"odinger equation with the
potential $V^{(1)}$ admits {\it SV} in the Cartesian coordinate system
$\omega_0=t,\ \omega_1=x_1,\ \omega_2=x_2$ only, while the one with
the potential $k_1x_1^2+V_2(x_2)$ separates in seven ($k_1<0$) or in
three ($k_1>0$) coordinate systems.

A complete list of coordinate systems providing {\it SV} in the
Schr\"odinger equations with the above given potentials takes two
dozens of pages. Therefore, we restrict ourself to considering the
Schr\"odinger equation with anisotropic harmonic oscillator potential
$V(x_1, x_2)= k_1x_1^2+k_2x_2^2$, $k_1\ne k_2$ and Coulomb potential
$V(x_1, x_2)= k_1(x_1^2+x_2^2)^{-1/2}$.

\section* {III. Separation of variables in the Schr\"odinger equation
  with the anisotropic harmonic oscillator and the Coulomb
  potentials.}

Here we will obtain all coordinate systems providing separability
of the Schr\"odinger equation  with the potential $V(x_1, x_2)=
k_1x_1^2+k_2x_2^2$

\begin{equation}
\imath u_t+u_{x_1x_1}+u_{x_2x_2}=(k_1x_1^2+k_2x_2^2)u.
\label{11}
\end{equation}

In the following, we consider the case $k_1\ne k_2$,
because otherwise Eq.(\ref{1}) is reduced to the free Schr\" odinger
equation (see the Remark 2) which has been studied in detail in
[1-3].

Explicit forms of the coordinate systems to be found depend
essentially on the signs of the parameters $k_1, \ k_2$. We
consider in detail the case, when $k_1<0, \ k_2>0$ (the cases $k_1>0,
\ k_2>0$ and $k_1<0, \ k_2<0$ are handled in an analogous way). It
means that Eq.(\ref{11}) can be written in the form

\begin{equation}
\imath u_t+u_{x_1x_1}+u_{x_2x_2}+{1\over 4}(a^2x_1^2-b^2x_2^2)u=0,
\label{12}
\end{equation}
where $a,\ b$ are arbitrary non-null real constants (the factor
${1\over 4}$ is introduced for further convenience).

As stated above to describe all coordinate systems $t, \ \omega
_1(t,\vec x), \ \omega _2(t,\vec x)$ providing separability of
Eq.(\ref{11}) one has to construct the general solution of system
(\ref{7}) with $V(x_1,x_2)= -{1\over 4}(a^2x_1^2-b^2x_2^2)$. The
general solution of equations 1--3 from (\ref{6}) splits into four
inequivalent classes listed in (\ref{7}).  Analysis shows that only
solutions belonging to the first class can satisfy an equation 4 from
(\ref{6}).

Substituting the expressions for $\omega _1, \ \omega _2, \ Q$
given by the  formulae 1 from (\ref{7}) into the equation 4 from
(\ref{6}) with $V(x_1,x_2)=-{1\over 4}(a^2x_1^2-b^2x_2^2)$ and
splitting with respect to $x_1, \ x_2$ one gets

\begin{eqnarray}
& &B_{01}(\omega _1)=\alpha _1\omega _1^2+\alpha _2\omega _1, \
B_{02}(\omega _2)=\beta _1\omega _2^2+\beta _2\omega _2,
\nonumber\\[4mm]
& &\biggl ({\dot A\over A}\biggr )^{^{\ds .}}-\biggl ({\dot A\over
  A}\biggr )^2 -4\alpha _1A^4+a^2=0,\label{13a} \\
& &\biggl ({\dot B\over B}\biggr )^{^{\ds .}}-\biggl ({\dot B\over
  B}\biggr )^2 -4\beta _1B^4-b^2=0,\label{13b} \\
& &\ddot \theta _1-2\dot \theta _1{\dot A\over
A}-2(2\alpha_1\theta _1+\alpha _2)A^4=0,\label{13c} \\
& &\ddot \theta _2-2\dot \theta _2{\dot B\over
B}-2(2\beta_1\theta _2+\beta _2)B^4=0.\label{13d}
\end{eqnarray}

Here $\alpha _1, \ \alpha _2, \ \beta _1, \ \beta _2$ are arbitrary
real constants.

Integration of the system of nonlinear {\it ODEs}
(\ref{13a})--(\ref{13d}) is carried out in the Appendix 1.
Substitution of the formulae (A.2), (A.4)-(A.6), (A.8)-(A.11) into the
corresponding expressions 1 from (\ref{7}) yields a complete list of
coordinate systems providing separability of the Schr\" odinger
equation (\ref{12}). These systems can be transformed to canonical
form if we use the Remark 3.

The invariance group of Eq.(\ref{12}) is generated by the following
basis operators \cite{bo2}:

\begin{eqnarray}
& &P_0=\partial _t, \quad I=u\partial _u, \quad M=\imath u\partial _u,
\quad Q_{\infty}=U(t,\vec x)\partial _u \nonumber\\
& &P_1=\cosh at \, \partial _{x_1}+\ds {\imath a\over 2}
(x_1\sinh at)u\partial _u,\nonumber\\
& &P_2=\cos bt \, \partial _{x_2}-\ds {\imath b\over 2}
(x_2\sin bt)u\partial _u,\label{14} \\
& &G_1=\sinh at \, \partial _{x_1}+\ds {\imath a\over 2}
(x_1\cosh at)u\partial _u,\nonumber\\
& &G_2=\sin bt \, \partial _{x_2}+\ds {\imath b\over 2}
(x_2\cos bt)u\partial _u,\nonumber
\end{eqnarray}
where $U(t, \vec x)$ is an arbitrary solution of Eq.(\ref{12}).

Using the finite transformations generated by the infinitesimal
operators (\ref{14}) and the Remark 3 we can choose in the formulae
(A.4)-(A.6), (A.8), (A.10), (A.11) $C_3=C_4=D_1=0, \ D_3=D_4=0,
\ C_2=D_2=1$. As a result, we come to the following assertion.
\vskip 2mm

\noindent
{\large \bf Theorem 1.} {\it The Schr\"odinger equation (\ref{12})
  admits {\it SV} in 21 inequivalent coordinate systems of the form

\begin{equation}
\omega _0=t, \quad \omega _1=\omega _1(t, \vec x), \quad
\omega _2=\omega _2(t, \vec x),
\label{15}
\end{equation}
where $\omega _1$ is given by one of the formulae from the first and $
\omega_2$ -- by one of the formulae from the second column of the
Table 1.}

\begin{center}{\it Table 1.}\end{center}
\vskip 2mm
\begin{center}
\begin{tabular}{|c||c|}\hline
 & \\
$\omega_1(t,\vec x)$ & $\omega_2(t,\vec x)$\\
 & \\ \hline
 & \\
$x_1\Bigl (\sinh a(t+C)\Bigl )^{-1}+
\alpha \Bigl (\sinh a(t+C)\Bigr )^{-2}$ & $x_2(\sin bt)^{-1}+\beta
(\sin bt)^{-2}$\\
$x_1\Bigl (\cosh a(t+C)\Bigl )^{-1}+
\alpha \Bigl (\cosh a(t+C)\Bigr )^{-2}$ & $x_2(\beta +\sin
2bt)^{-1/2}$\\
$x_1\exp (\pm at)+ \alpha \exp (\pm 4at)$ & $x_2$ \\
$x_1\Bigl (\alpha +\sinh 2a(t+C)\Bigr )^{-1/2}$ & \\
$x_1\Bigl (\alpha +\cosh 2a(t+C)\Bigr )^{-1/2}$ & \\
$x_1\Bigl (\alpha +\exp (\pm 2at)\Bigr )^{-1/2}$ & \\
$x_1$ & \\
 & \\ \hline
\end{tabular}
\end{center}
\vskip 2mm

{\it Here $C, \ \alpha , \ \beta $ are arbitrary real
constants.}

There is no necessity to consider specially the case when in
(\ref{11}) $k_1>0, \ k_2<0$, since such an equation by the change of
independent variables $u(t,x_1,x_2)\to u(t,x_2,x_1)$ is reduced to
Eq.(\ref{12}).

Below we adduce without proof the assertions describing coordinate
systems providing {\it SV} in Eq.(\ref{11}) with $k_1<0, \ k_2<0$ and
$k_1>0, \ k_2>0$.
\vskip 2mm

\noindent
{\large \bf Theorem 2.} {\it The Schr\" odinger equation

\begin{equation}
\imath u_t+u_{x_1x_1}+u_{x_2x_2}+
{1\over 4}(a^2x_1^2+b^2x_2^2)u=0
\label{18}
\end{equation}
with $a^2\ne 4b^2$ admits {\it SV} in 49 inequivalent coordinate
systems of the form (\ref{15}), where $\omega_1$ is given by one of
the formulae from the first and $\omega_2$ -- by one of the formulae
from the second column of the Table 2. Provided $a^2=4b^2$ one more
coordinate system should be included into the above list, namely}

\begin{equation}
\omega_0=t, \quad \omega_1^2-\omega_2^2 =2x_1, \quad \omega
_1\omega_2=x_2.
\label{19}
\end{equation}

\begin{center}{\it Table 2.}\end{center}
\vskip 2mm

\begin{center}
\begin{tabular}{|c||c|}\hline
 & \\
$\omega_1(t,\vec x)$ & $\omega_2(t,\vec x)$\\
 & \\ \hline
 & \\
$x_1\Bigl (\sinh a(t+C)\Bigl )^{-1}+
\alpha \Bigl (\sinh a(t+C)\Bigr )^{-2}$ & $x_2(\sinh bt)^{-1}+\beta
(\sinh bt)^{-2} $\\
$x_1\Bigl (\cosh a(t+C)\Bigl )^{-1}+
\alpha \Bigl (\cosh a(t+C)\Bigr )^{-2}$ & $x_2(\cosh bt)^{-1}+\beta
(\cosh bt)^{-2}$\\
$x_1\exp (\pm at)+
\alpha \exp (\pm 4at)$ & $x_2\exp (\pm bt)+\beta \exp (\pm  4bt)$ \\
$x_1\Bigl (\alpha +\sinh 2a(t+C)\Bigr )^{-1/2}$ &$x_2(\beta +\sinh
2bt)^{-1/2}$ \\
$x_1\Bigl (\alpha +\cosh 2a(t+C)\Bigr )^{-1/2}$ &$x_2(\beta +\cosh
2bt)^{-1/2}$ \\
$x_1\Bigl (\alpha +\exp (\pm 2at)\Bigr )^{-1/2}$ &$x_2\Bigl (\beta
+\exp (\pm 2bt)\Bigr )^{-1/2}$ \\
$x_1$ & $x_2$\\
 & \\ \hline
\end{tabular}
\end{center}
\vskip 2mm

Here $C,\ \alpha,\ \beta$ are arbitrary constants.
\vskip 2mm

\noindent
{\large \bf Theorem 3.} {\it The Schr\" odinger equation

\begin{equation}
\imath u_t+u_{x_1x_1}+u_{x_2x_2}-{1\over
4}(a^2x_1^2+b^2x_2^2)u=0
\label{20}
\end{equation}
with $a^2\ne 4b^2$ admits {\it SV} in 9 inequivalent coordinate systems of
the form (\ref{15}), where $\omega _1$ is given by one of the formulae
from the first and $\omega_2$ -- by one of the formulae from the
second column of the Table 3. Provided $a^2=4b^2$, the above list
should be supplemented by the coordinate system (\ref{19}).}

\begin{center}{\it Table 3.}\end{center}
\vskip 2mm

\begin{center}
\begin{tabular}{|c||c|}\hline
 & \\
$\omega_1(t,\vec x)$ & $\omega_2(t,\vec x)$\\
 & \\ \hline
 & \\
$x_1\Bigl (\sin a(t+C)\Bigl )^{-1}+
\alpha \Bigl (\sin a(t+C)\Bigr )^{-2}$ & $x_2(\sin bt)^{-1}+\beta
(\sin bt)^{-2}$\\
$x_1\Bigl (\beta +\sin 2a(t+C)\Bigl )^{-1/2}$ & $x_2(\beta +\sin
2bt)^{-1/2}$\\
$x_1$ & $x_2$ \\
 & \\ \hline
\end{tabular}
\end{center}
\vskip 2mm

Here $C,\ \alpha,\ \beta$ are arbitrary constants.

{\it Remark 4.} If we consider (1) as an equation for a complex-valued
function $u$ of three complex variables $t,\ x_1,\ x_2$, then the cases
considered in the Theorems 1--3 are equivalent. Really, replacing, when
necessary, $a$ with $\imath a$ and $b$ by $\imath b$ we can always
reduce Eqs.(\ref{12}), (\ref{18}) to the form (\ref{20}). It means that
coordinate systems presented in the Tables 1, 2 are complex equivalent
to those listed in the Table 3. But if $u$ is a complex-valued
function of real variables $t,\ x_1,\ x_2$ it is not the case.
\vskip 2mm

\noindent
{\large \bf Theorem 4.} {\it The Schr\" odinger equation with the
  Coulomb potential

$$
\imath u_t+u_{x_1x_1}+u_{x_2x_2}-
k_1(x_1^2+x_2^2)^{-1/2}u=0
$$
admits {\it SV} in two coordinate systems (\ref{12*}), (\ref{19}).}

It is important to note that explicit forms of coordinate systems
providing separability of Eqs.(\ref{12}), (\ref{18}), (\ref{20})
depend essentially on the parameters $a, \; b$ contained in the
potential $V(x_1,x_2)$.  It means that the free Schr\"odinger equation
($V=0$) does not admit {\it SV} in such coordinate systems.
Consequently, they are essentially new.

\section* {IV. Conclusion}

In the present paper we have studied the case when the
Schr\"odinger equation (\ref{1}) separates into one first-order
and two second-order {\it ODEs}. It is not difficult to prove
that there are no functions $Q(t,\vec x), \ \omega _{\mu}(t,\vec x),
\ \mu =\overline{0,2}$ such that the Ansatz

$$
u=Q(t,\vec x)\varphi _0(\omega _0(t,\vec x))
\varphi _1\Bigl (\omega _1(t,\vec x)
\Bigr)\varphi_2\Bigl (\omega _2(t,\vec x)\Bigr )
$$
separates Eq.(\ref{1}) into three second-order {\it ODEs} (see
Appendix 2). Nevertheless, there exists a possibility for Eq.(\ref{1})
to be separated into two first-order and one second-order {\it ODEs}
or into three first-order {\it ODEs}. This is a probable source of new
potentials and new coordinate systems providing separability of the
Schr\"odinger equation. It should be said that separation of
the two-dimensional wave equation
$$
u_{tt}-u_{xx}=V(x)u
$$
into one first-order and one second-order {\it ODEs} gives no new
potentials as compared with separation of it into two second-order
{\it ODEs}. But for some already known potentials new coordinate
system providing separability of the above equation are obtained
\cite{zrfu}.

Let us briefly analyze a connection between separability of
Eq.(\ref{1}) and its symmetry properties. It is well-known that each
solution of the free Schr\" odinger equation with separated variables
is a common eigenfunction of two mutually commuting second-order
symmetry operators of the said equation \cite{bokm,mi}. And what is
more, separation constants $\lambda _1, \ \lambda _2$ are eigenvalues
of these symmetry operators.

We will establish that the same assertion holds for the Schr\" odinger
equation (\ref{1}). Let us make in Eq.(\ref{1}) the following change
of variables:

\begin{equation}
u=Q(t,\vec x)U\Bigl (t,\omega _1(t,\vec x),
\omega _2(t,\vec x)\Bigr ),
\label{21}
\end{equation}
where $(Q, \; \omega _1, \; \omega _2)$ is an arbitrary solution
of the system of {\it PDEs} (\ref{6}).

Substituting the expression (\ref{21}) into (\ref{1}) and taking into
account equations (\ref{6}) we get

\begin{equation}
Q\biggl (\imath U_t+\Bigl (U_{\omega _1\omega _1}-
B_{01}(\omega _1)U\Bigr )\omega _{1x_a}\omega _{1x_a}+
\Bigl (U_{\omega _2\omega _2}-B_{02}(\omega _2)U\Bigr )
\omega _{2x_a}\omega _{2x_a}\biggr )=0.
\label{22}
\end{equation}

Resolving equations 2 from the system (\ref{6}) with respect to
$\omega_{1x_a}\omega_{1x_a}$ and $\omega_{2x_a}\omega_{2x_a}$
we have

\begin{eqnarray*}
\omega_{1x_a}\omega_{1x_a}&=&{1\over \delta}\Bigl (R_2(t)
B_{21}(\omega _2)-R_1(t)B_{22}(\omega _2)\Bigr ), \\
\omega_{2x_a}\omega_{2x_a}&=&{1\over \delta}\Bigl (R_1(t)
B_{12}(\omega _1)-R_2(t)B_{11}(\omega _1)\Bigr ).
\end{eqnarray*}
where $\delta =B_{11}(\omega _1)B_{22}(\omega _2)-
B_{12}(\omega _1)B_{21}(\omega _2)$ ($\delta \ne 0$ by force of
the condition (\ref{3b})).

Substitution of the above equalities into Eq.(\ref{22}) with subsequent
division by $Q\ne 0$ yields the following {\it PDE}:

\begin{equation}
\begin{array}{rcl}
\imath U_t&+&\ds {R_1(t)\over \delta }\biggl (B_{12}(\omega _1)
\Bigl (U_{\omega _2\omega _2} - B_{02}(\omega _2)U\Bigr )-
B_{22}(\omega _2)\Bigl (U_{\omega _1\omega _1}-
B_{01}(\omega _1)U\Bigr )\biggr ) \\
&+&\ds {R_2(t)\over \delta }\biggl (B_{21}(\omega _2)
\Bigl (U_{\omega _1\omega _1}-B_{01}(\omega _1)U\Bigr )-
B_{11}(\omega _1)\Bigl (U_{\omega _2\omega _2}-
B_{02}(\omega _2)U\Bigr )\biggr )=0.
\end{array}
\label{23}
\end{equation}

Thus, in the new coordinates $t, \ \omega _1, \ \omega _2, \ U(t,\omega _1,
\omega_2)$ Eq.(\ref{1}) takes the form (\ref{23}).

By direct (and very cumbersome) computation one can check
that the following second-order differential operators:

\begin{eqnarray*}
X_1&=&
\ds {B_{22}(\omega _2)\over \delta }\Bigl (\partial _{\omega _1}^2-
B_{01}(\omega _1)\Bigr )-\ds {B_{12}(\omega _1)\over \delta }
\Bigl (\partial _{\omega _2}^2 - B_{02}(\omega _2)\Bigr ),\\
X_2&=&
-\ds {B_{21}(\omega _2)\over \delta }\Bigl (\partial _{\omega _1}^2-
B_{01}(\omega _1)\Bigr )+\ds {B_{11}(\omega _1)\over \delta }
\Bigl (\partial _{\omega _2}^2 - B_{02}(\omega _2)\Bigr ),
\end{eqnarray*}
commute under arbitrary $B_{0a}, \ B_{ab}, \ a,b=1,2$, i.e.

\begin{equation}
\Bigl [X_1, \ X_2\Bigr ]\equiv X_1X_2-X_2X_1=0.
\label{24}
\end{equation}

After being rewritten in terms of the operators $X_1$, \ $X_2$
Eq.(\ref{23}) reads

$$
\Bigl (\imath \partial _t-R_1(t)X_1-R_2(t)X_2\Bigr )U=0.
$$

Since the relations

\begin{equation}
\Bigl [\imath \partial_t-R_1(t)X_1-R_2(t)X_2, \ X_a\Bigr ]=0, \ a=1,2
\label{25}
\end{equation}
hold, operators $X_1$, \ $X_2$ are mutually commuting symmetry
operators of Eq.(\ref{23}). Furthermore,  solution of Eq.(\ref{23})
with separated variables $U=\varphi _0(t)\varphi _1(\omega _1)
\varphi _2(\omega _2)$ satisfies the identities

\begin{equation}
X_aU=\lambda _aU, \ a=1,2.
\label{26}
\end{equation}

Consequently, if we designate by $X_1^{\prime }$, \ $X_2^{\prime }$
the operators $X_1$, \ $X_2$ written in the initial variables
$t, \ \vec x, \ u$, then we get from (\ref{24})--(\ref{26}) the following
equalities:

\begin{eqnarray*}
& &\Bigl [\imath \partial _t+\triangle -V(x_1,x_2), \
X_a^{\prime }\Bigr ]=0, \ a=1,2, \\
& &\Bigl [X_1^{\prime }, \ X_2^{\prime }\Bigr ]=0, \\
& &X_a^{\prime }u=\lambda _au, \ a=1,2.
\end{eqnarray*}
where $u=Q(t,\vec x)\varphi _0(t)\varphi _1(\omega _1)
\varphi _2(\omega _2)$.

It means that each solution with separated variables is a
common eigenfunction of two mutually commuting symmetry
operators $X_1^{\prime }, \ X_2^{\prime }$ of the Schr\" odinger
equation (\ref{1}), separation constants $\lambda _1, \ \lambda _2$
being their eigenvalues.

Detailed study of the said operators as well as analysis of separated
{\it ODEs} for functions $\varphi _{\mu }, \ \mu =\overline{0,2}$ (in
the way as it is done for the free Schr\"odinger equation in
\cite{bokm,mi}) is in progress and will be a topic of our future
publications.

\section* {V. Acknowledgments}

When the paper was in the last stage of preparation, one
of the authors (R.Zhdanov) was supported by the Alexander
von Humboldt Foundation. Taking an opportunity he wants
to express his gratitude to Director of Arnold Sommerfeld
Institute for Mathematical Physics Professor H.-D.Doebner
for hospitality.

\bigskip
\bigskip
\appendix {\bf Appendix 1. Integration of nonlinear ODEs
(\ref{13a})--(\ref{13d}).}
\medskip

Evidently, equations (\ref{13a})--(\ref{13d}) can be rewritten in the
following unified form:

$$
\biggl (\ds {\dot y\over y}\biggr )^{^{\ds .}}-\biggl
(\ds {\dot y\over y}\biggr )^2-4\alpha y^4=k,\quad
\ddot z-2\dot z\ds {\dot y\over
y}-2(2\alpha z+\beta )y^4=0.
\eqno(A.1)
$$

Provided $k=-a^2<0$, system (A.1) coincides with equations
(\ref{13a}), (\ref{13c}) and under $k=b^2>0$ -- with equations
(\ref{13b}), (\ref{13d}).

First of all, we note that the function $z=z(t)$ is determined up
to addition of an arbitrary constant. Really, the coordinate
functions $\omega _a$ have the following structure:

$$
\omega _a=yx_a+z, \ a=1,2.
$$

But the coordinate system $t, \ \omega _1, \ \omega _2$ is equivalent
to the coordinate system $t, \ \omega _1+C_1, \ \omega _2+C_2, \
C_a\in R^1$.  Hence it follows that the function $z(t)$ is equivalent
to the function $z(t)+C$ with arbitrary real constant $C$.
Consequently, provided $\alpha \ne 0$, we can choose in (A.1) $\beta
=0$.

{\it The case 1.} $\alpha =0$.

On making in (A.1) the change of variables

$$w={\dot y\over y}, \quad v={z\over y}\eqno(A.2)$$
we get

$$\dot w=w^2+k, \quad \ddot v+kv=2\beta y^3.\eqno(A.3)$$

First, we consider the case $k=-a^2<0$. Then the general solution
of the first equation from (A.3) is given by one of the formulae

$$w=-a\coth a(t+C_1), \quad  w=-a\tanh a(t+C_1), \quad w=\pm a, \quad
C_1\in R^1,$$
whence

$$y=C_2\sinh ^{-1} a(t+C_1), \quad y=C_2\cosh ^{-1} a(t+C_1),\quad
y=C _2\exp (\pm at), \ C_2\in R^1.
\eqno(A.4)
$$

The second equation of system (A.3) is a linear inhomogeneous {\it ODE}.
Its general solution after being substituted into (A.2) yields
the following expression for $z(t)$:

$$
\begin{array}{l}
(C_3\cosh at+C_4\sinh at)\sinh ^{-1}a(t+C_1)
+\ds {\beta C_2^4\over a^2}\sinh ^{-2} a(t+C_1), \\[2mm]
(C_3\cosh at+C_4\sinh at)\cosh ^{-1}a(t+C_1)
+\ds {\beta C_2^4\over a^2}\cosh ^{-2} a(t+C_1), \\[2mm]
(C_3\cosh at+C_4\sinh at)\exp (\pm at)
+\ds {\beta C _2^4\over 4a^2}\exp (\pm 4at), \quad
C_3,C_4\subset R^1.
\end{array}
\eqno(A.5)
$$

The case $k=b^2>0$ is treated in an analogous way, the general
solution of (A.1) being given by the formulae

$$\begin{array}{l}
y=D_2\sin ^{-1}b(t+D_1),\\[2mm]
z=(D_3\cos bt+D_4\sin bt)\sin ^{-1}b(t+D_1)
+\ds {\beta D_2^4\over b^2}\sin ^{-2} b(t+D_1),
\end{array}\eqno(A.6)$$
where $D_1, \ D_2, \ D_3, \ D_4$ are arbitrary real constants.

{\it The case 2.} $\alpha \ne 0, \ \beta =0$.

On making in (A.1) the change of variables

$$y=\exp w, \quad v={z\over y}$$
we have

$$\ddot w-\dot w^2=k+\alpha \exp 4w, \quad  \ddot v +kv=0.\eqno(A.7)$$

The first {\it ODE} from (A.7) is reduced to the first-order linear {\it ODE}

$${1\over 2}{dp(w)\over dw}-p(w)=k+\alpha \exp 4w$$
by the substitution $\dot w=\Bigl (p(w)\Bigr )^{1/2}$, whence

$$p(w)=\alpha \exp 4w +\gamma \exp 2w -k, \ \gamma \in R^1.$$

Equation $\dot w=\Bigl(p(w)\Bigr)^{1/2}$ has a singular solution $w=C=const$
such that $p(C)=0$. If $\dot w \ne 0$, then integrating the equation
$\dot w=p(w)$ and returning to the initial variable $y$ we get

$$
\int\limits _{ }^{\displaystyle y(t)}{d\tau \over \tau(\alpha \tau
  ^4+\gamma \tau ^2-k)^{1/2}}=t +C_1.
$$

Taking the integral in the left-hand side of the above equality we
obtain the general solution of the first {\it ODE} from (A.1). It is given
by the following formulae:

\noindent\underline {under $k=-a^2<0$}

$$\begin {array} {l}y=C_2\Bigl (\alpha +\sinh 2a(t+C_1)\Bigr )^{-1/2},\\
y=C_2\Bigl (\alpha +\cosh 2a(t+C_1)\Bigr )^{-1/2},\\
y=C_2\Bigl (\alpha +\exp (\pm 2at)\Bigr )^{-1/2},\\
\end{array}\eqno(A.8)$$

\noindent\underline {under $k=b^2>0$}

$$y=D_2\Bigl (\alpha +\sin 2b(t+D_1)\Bigr )^{-1/2}.\eqno(A.9)$$

Here $C_1, \ C_2, \ D_1, \ D_2$ are arbitrary real constants.

Integrating the second {\it ODE} from (A.7) and returning to the initial
variable $z$ we have

\noindent\underline {under $k=-a^2<0$}

$$z=y(t)(C_3\cosh at+C_4\sinh at)\eqno(A.10)$$

\noindent\underline {under $k=b^2>0$}

$$z=y(t)(D_3\cos bt+D_4\sin bt),\eqno(A.11)$$
where $C_3, \ C_4, \ D_3, \ D_4$ are arbitrary real constants.

Thus, we have constructed the general solution of the system
of nonlinear {\it ODEs} (A.1) which is given by the formulae (A.5)--(A.11).

\bigskip

\appendix {\bf Appendix 2. Separation of Eq.(\ref{1}) into three
  second-order ODEs.}
\medskip

Suppose that there exists an Ansatz

$$u=Q(t,\vec x)\varphi _0(\omega _0(t,\vec x))
\varphi _1\Bigl (\omega _1(t,\vec x)
\Bigr)\varphi_2\Bigl (\omega _2(t,\vec x)\Bigr )\eqno(A.12)$$
which separates the Schr\"odinger equation into three second-order
{\it ODEs}

$$
\begin{array}{rcl}
\ds {d^2\varphi _0\over d\omega _0^2}&=&U_0(\omega _0,
\varphi _0, \ds {d\varphi _0\over d\omega _0};\lambda _1,\lambda _2),
\\[2mm] \ds {d^2\varphi _1\over d\omega _1^2}&=&U_1(\omega _1,
\varphi _1, \ds {d\varphi _1\over d\omega _1};\lambda _1,\lambda _2),
\\[2mm] \ds {d^2\varphi _2\over d\omega _2^2}&=&U_2(\omega _2,
\varphi _2, \ds {d\varphi _2\over d\omega _2};\lambda _1,\lambda _2)
\end{array}\eqno(A.13)
$$
according to the Definition 1.

Substituting the Ansatz (A.12) into Eq.(\ref{1}) and excluding the second
derivatives ${d^2\varphi _{\mu }\over d\omega _{\mu }^2}$,
$\ \mu =\overline {0,2}$ according to Eqs.(A.13) we get

\begin{eqnarray*}
& &\imath (Q_t\varphi _0\varphi _1\varphi _2
+Q\omega _{0t}\dot \varphi _0\varphi _1\varphi _2
+Q\omega _{1t}\varphi _0\dot \varphi _1\varphi _2
+Q\omega _{2t}\varphi _0\varphi _1\dot \varphi _2)
+(\triangle Q)\varphi _0\varphi _1\varphi _2\\
& &\quad +2Q _{x_a}\omega _{0x_a}\dot \varphi _0\varphi _1\varphi _2
+2Q _{x_a}\omega _{1x_a}\varphi _0\dot \varphi _1\varphi _2
+2Q _{x_a}\omega _{2x_a}\varphi _0\varphi _1\dot \varphi _2
+Q\Bigl ((\triangle \omega _0)\dot \varphi _0\varphi _1\varphi _2\\
& &\quad +(\triangle \omega _1)\varphi _0\dot \varphi _1\varphi _2
+(\triangle \omega _2)\varphi _0\varphi _1\dot \varphi _2
+\omega _{0x_a}\omega_ {0x_a}U _0\varphi _1\varphi _2
+\omega _{1x_a}\omega_ {1x_a}\varphi _0U_1\varphi _2\\
& &\quad +\omega _{2x_a}\omega_ {2x_a}\varphi _0\varphi _1U_2
+2\omega _{0x_a}\omega _{1x_a}\dot \varphi _0\dot \varphi _1\varphi _2
+2\omega _{0x_a}\omega _{2x_a}\dot \varphi _0\varphi _1\dot \varphi _2
+2\omega _{1x_a}\omega _{2x_a}\varphi _0\dot \varphi _1\dot \varphi
_2\Bigr )\\
& &\quad =VQ\varphi _0\varphi _1\varphi _2.
\end{eqnarray*}

Splitting the above equality with respect to $\dot \varphi _0
\dot \varphi _1, \ \dot \varphi _0\dot \varphi _2, \ \dot \varphi _1
\dot \varphi _2$ we obtain the equalities:

$$
\omega _{0x_a}\omega _{1x_a}=0,\quad
\omega _{0x_a}\omega _{2x_a}=0,\quad
\omega _{1x_a}\omega _{2x_a}=0.\eqno(A.14)
$$

Since the functions $\omega _{\mu }, \ \mu =\overline {0,2}$ are
real-valued, equalities (A.14) mean that there are three real
two-component vectors which are mutually orthogonal. This is possible
only if one of them is a null-vector. Without loss of generality
we may suppose that $(\omega _{0x_1}, \; \omega _{0x_2})=(0,0)$,
whence $\omega _0=f(t)\sim t$.

Consequently, Ansatz (A.12) necessarily takes the form (\ref{4}).  But
Ansatz (\ref{4}) can not separate Eq.(\ref{1}) into three second-order
{\it ODEs}, since it contains no second-order derivative with respect
to $t$.

Thus, we have proved that the Schr\"odinger equation (\ref{1}) is not
separable into three second-order {\it ODEs}.


\begin{thebibliography}{99}
\bibitem{bo1} C.Boyer, SIAM J.Math.Anal. {\bf 7}, 230 (1976).

\bibitem{bokm} C.Boyer, E.Kalnins and W.Miller, J.Math.Phys. {\bf
    16}, 499 (1975).

\bibitem{mi} W.Miller, Symmetry and Separation of Variables
  (Addison-Wesley, Massachusetts, 1977).

\bibitem{sha} V.N.Shapovalov and N.B.Sukhomlin, Izvestiya
  Vuzov,Fizika N 12, 268 (1974).

\bibitem{niof} A.G.Nikitin, S.P.Onufriychuk and W.I.Fushchych,
   Teoret.Matem.Fizika {\bf 12}, 268 (1992).

\bibitem{fuzr} W.I.Fushchych, R.Z.Zhdanov and I.V.Revenko,
  Proc.Ukrain.Acad.Sci. N 1, 27 (1993).

\bibitem{fu} W.I.Fushchych, in: Symmetry in Mathematical Physics
  Problems (Inst.of Math., Kiev, 1981), p.6.

\bibitem{ko} T.H.Koornwinder, Lect.Notes in Math. {\bf 810}, 240
  (1980).

\bibitem{zrfu} R.Z.Zhdanov, I.V.Revenko and W.I.Fushchych, J.Phys.A:
  Math.Gen. {\bf 26}, 5959 (1993).

\bibitem{zh} R.Z.Zhdanov, J.Phys.A: Math.Gen. {\bf 27}, L291 (1994).

\bibitem{bo2} C.Boyer Helv.Phys.Acta {\bf 47}, 589 (1974).

\bibitem{ni} U.Niederer Helv.Phys.Acta {\bf 45}, 802 (1972).
\end{thebibliography}
\end{document}